\documentclass[10pt,journal]{IEEEtran}
\usepackage{cite}
\usepackage{amssymb}
\usepackage{bm}
\usepackage{graphicx,subfigure}
\usepackage[T1]{fontenc}
\usepackage{textcomp}

\begin{document}

\title{Toward the Internet of No Things: The Role of O2O Communications and Extended Reality\thanks{This work was supported by NSERC Discovery Grant 2016-04521 and the FRQNT B2X doctoral scholarship programme.}}

\author{Martin Maier and Amin Ebrahimzadeh\thanks{M. Maier and A. Ebrahimzadeh are with the Optical Zeitgeist Laboratory, Institut National de la Recherche Scientifique (INRS), Montr\'eal, QC, H5A 1K6, Canada. Email: \texttt{maier@ieee.org}}}
                
\maketitle
        
\begin{abstract}
Future fully interconnected virtual reality (VR) systems and the Tactile Internet diminish the boundary between virtual (online) and real (offline) worlds, while extending the digital and physical capabilities of humans via edge computing and teleoperated robots, respectively. In this paper, we focus on the \emph{Internet of No Things} as an extension of immersive VR from virtual to real environments, where human-intended Internet services---either digital or physical---appear when needed and disappear when not needed. We first introduce the concept of integrated online-to-offline (O2O) communications, which treats online and offline channels as complementary to bridge the virtual and physical worlds and provide O2O multichannel experiences. We then elaborate on the emerging extended reality (XR), which brings the different forms of virtual/augmented/mixed reality together to realize the entire reality-virtuality continuum and, more importantly, supports human-machine interaction as envisioned by the Tactile Internet, while posing challenges to conventional handhelds, e.g., smartphones. Building on the so-called \emph{invisible-to-visible (I2V)} technology concept, we present our \emph{extrasensory perception network (ESPN)} and investigate how O2O communications and XR can be combined for the nonlocal extension of human ``sixth-sense" experiences in space and time. We conclude by putting our ideas in perspective of the 6G vision.
\end{abstract}
        
\section{Introduction}
\label{sec:intro}
The International Telecommunication Union (ITU) has divided 5G mobile network services into the following three categories: ($i$) enhanced mobile broadband (eMBB), ($ii$) ultra-reliable low-latency communications (URLLC), and ($iii$) massive machine type communications (mMTC). The latter one is also referred to as machine-to-machine (M2M) communications in the context of the Internet of Things (IoT). The IoT without any human involvement in its underlying M2M communications is useful for the automation of industrial and other machine-centric processes while keeping the human largely out of the loop. Conversely, some of the most interesting 5G applications designed to keep the human in the loop sit at the crossroads between eMBB and URLLC. Among others, \emph{virtual reality (VR)} is expected to be one of the killer applications in 5G networks~\cite{EPBD18}.

According to~\cite{BBMD17}, VR systems will undergo the following three evolutionary stages. The first evolutionary stage includes current VR systems that require a wired connection to a PC or portable device because current 4G or even pre-5G wireless systems cannot cope with the massive amount of bandwidth and latency requirements of VR. The PC or portable device in turn is connected to the central cloud and the Internet via backhaul links. At the second evolutionary stage, VR devices are wirelessly connected to a fog/edge server located at the base station for local computation and caching. The third and final evolutionary stage envisions ideal (fully interconnected) VR systems, where no distinction between real and virtual worlds are made in human perception. In addition, according to~\cite{BBMD17}, the growing number of drones, robots, and self-driving vehicles (to be discussed in more detail in Section~\ref{sec:I2V}) will take cameras to places humans could never imagine reaching. 

Another interesting 5G application that involves the inherent human-in-the-loop nature of human-machine interaction is the so-called \emph{Tactile Internet}, which allows for the tactile steering and control of not only virtual but also real objects, e.g., teleoperated robots, as well as processes. The emerging IEEE P1918.1 standards working group develops a generic Tactile Internet reference architecture based on several key principles. Among others, the key principles envision to support local area as well as wide area connectivity through wireless (e.g., cellular, WiFi) or hybrid wireless/wired networking and to leverage computing resources from cloud variants at the edge of the network. Recently, in~\cite{MaEC18}, we showed that the human-centric design approach of the Tactile Internet helps extend the capabilities of humans through the Internet by supporting them in the coordination of their physical and digital co-activities with robots and software agents by means of artificial intelligence (AI) enhanced multi-access edge computing (MEC). 

The above discussion shows that future fully interconnected VR systems and the future Tactile Internet seem to evolve towards common design goals. Most notably, the boundary between virtual (i.e., \emph{online}) and real (i.e., \emph{offline}) worlds is to become increasingly imperceptible while both \emph{digital and physical capabilities of humans are to be extended} via edge computing variants, ideally with embedded AI capabilities. In fact, according to the inaugural report ``Artificial Intelligence and Life in 2030" of Stanford University's recently launched One Hundred Year Study on Artificial Intelligence (AI100), an increasing focus on developing systems that are human-aware is expected over the next 10-15 years. Clearly, such a shift of research focus represents an evolutionary leap from today's IoT without any human involvement in its underlying M2M communications. 

An interesting practical extension of immersive VR from virtual to a real environment is the so-called Naked world vision that aims at paving the way to an \emph{Internet of No Things} by offering all kinds of human-intended services without owning or carrying any type of computing or storage devices~\cite{AKLY18}. The term Internet of No Things was coined by Demos Helsinki founder Roope Mokka in 2015. The term nicely resonates with Eric Schmidt's famous statement at the 2015 World Economic Forum that ``the Internet will disappear" given that there will be so many things that we are wearing and interacting with that we won't even sense the Internet, though it will be part of our presence all the time. The Naked world envisions Internet services to appear from the surrounding environment when needed and disappear when not needed. Similar to the aforementioned evolution of today's VR systems, the transition from the current gadgets-based Internet to the Internet of No Things is divided into three phases that starts from \emph{bearables} (e.g., smartphone), moves towards \emph{wearables} (e.g., Google and Levi's smart jacket), and then finally progresses to the last phase of so-called \emph{nearables}. Nearables denote nearby surroundings or environments with embedded computing/storage technologies and service provisioning mechanisms that are intelligent enough to learn and react according to user context and history in order to provide user-intended services. According to~\cite{AKLY18}, their successful deployment is challenging not only from a technological point of view but also from a business and social mindset perspective due to the required user acceptability and trust.

In this paper, we focus on two emerging technologies that hold great promise to help realize the Internet of No Things by achieving the aforementioned common design goals of interconnected VR systems and the Tactile Internet. In Section~\ref{sec:O2O}, we introduce the concept of \emph{online-to-offline (O2O) communications} and review recent progress on how it may be used to tie both online and offline worlds closer together while opening new business opportunities and fostering user acceptability and trust. Section~\ref{sec:XR} elaborates on the reality-virtuality continuum that comprises various types of \emph{extended reality (XR)}, including not only VR but also augmented and mixed reality. In Section~\ref{sec:I2V}, we elaborate on how O2O and XR technologies can be leveraged toward realizing the Internet of No Things based on emerging \emph{invisible-to-visible} technologies. Finally, in Section~\ref{sec:6G}, we put our presented ideas in perspective of the future 6G vision and draw our conclusions in Section~\ref{sec:conclusions}.

\section{Online-to-Offline (O2O) Communications}
\label{sec:O2O}
The term O2O was first coined by Alex Rampell in 2010 after witnessing that the links between online and physical commerce are becoming stronger with the growth of local commerce on the Web such as Groupon. According to Rampell, the key to O2O is that it serves as a \emph{discovery mechanism} for finding consumers online, bringing them into real-world stores, and delivering social experiences like restaurants, given that most of the disposable income is spent locally offline. Despite the fact that millenials spend on average 7.5 hours a day online, it is interesting to note that 70\% of them prefer shopping in conventional brick-and-mortar stores because they want to have an experience that allows them to hold, touch, and in some cases try on products. Thus, it comes as no surprise that according to recent studies only 8.3\% of U.S. retail sales are done online while more than 80\% of them will still happen inside physical stores in 2020~\cite{Oren18}.

With the growing smartphone market, various attempts have been made to combine the merits of the diversity and affordability of online activities and the credibility and immediacy of offline activities, resulting in an expanded range of \emph{O2O services} beyond the initial concept of O2O commerce~\cite{KiCB17}. In the United States, O2O services started with lodging (AirBnB), transportation (Uber), and others such as neighborhood food delivery (e.g., GrubHub). An interesting example is Amazon's newest online-inspired Amazon 4-star brick-and-mortar store concept, which opened in New York City just recently on September 27, 2018. Amazon 4-star is a physical store where everything for sale is rated 4 stars and above, is a top seller, or is new and trending on the website Amazon.com. The concept is a clever example of experiential retail to fill customers with confidence in that everything they buy won't disappoint and inspire frequent visits due to dynamic assortment features such as local trends. Given that every product in the store has two different prices for Amazon Prime members and non-members, the concept also helps promote and drive Prime membership. In China, among other countries, QR codes have been widely used to display O2O services to customers, e.g., Tencent's so-called Weshopping service for Valentine's Day that directly orders the item a customer wants to buy online by merely scanning the QR code. Note that QR codes widen the notion of O2O in that they provide offline-to-online (rather than online-to-offline) services by connecting a customer scanning the QR code on a physical product with related online services. Thus, in general, the term O2O involves both online-to-offline and offline-to-online services to bridge the physical and virtual worlds and thereby provide integrated online-offline/offline-online experiences. 

\begin{figure}[t]
\centering
\includegraphics[width=.5\textwidth]{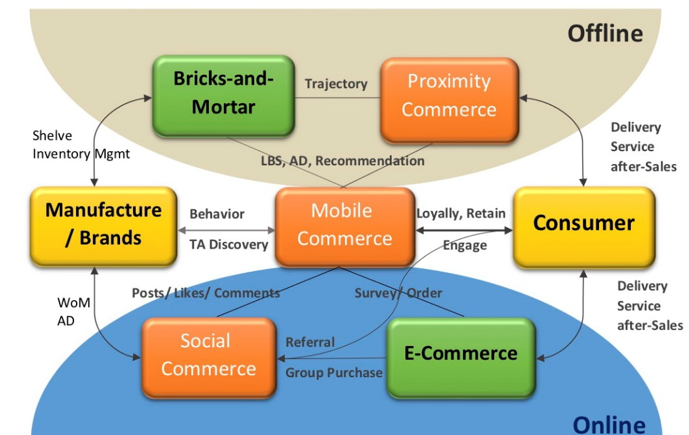}
\caption{Online-to-offline and offline-to-online (O2O) service model~\cite{TWLC15}. }
\label{fig1}
\end{figure}
\begin{table*}[t]
\centering
\caption{Online vs. offline advantages and disadvantages}
\label{tab:O2O}
\begin{tabular}{|c||l|l|}
\hline 
& \textbf{Online} & \textbf{Offline}  \\
\hline
\hline
\textbf{Advantages} & Low costs & Higher customer trust \\
\hline
& Zero inventory & Deeper customer experience \\
\hline
& Low investment risk & Instant gratification \\
\hline
& More information for customer & Personal face-to-face service \\
\hline
& One-click order possibility & Multi-modal physical interaction \\
\hline
& Easy scalability & Localized service \\
\hline
\hline
\textbf{Disadvantages} & Lower customer loyalty & Higher inventory costs \\
\hline
& Credit problems & Higher training costs \\
\hline
& Limited logistic service & Personnel quality demands \\
\hline
& Lack of physical interaction & Subject to waiting time in line-ups \\
\hline
& Shipping and handling & Product distribution and warehousing \\
\hline
\end{tabular}
\end{table*}
In~\cite{TWLC15}, the authors developed an \emph{O2O service model} for integrating online-to-offline and offline-to-online services. As shown in Fig.~\ref{fig1}, the proposed O2O service model relies on the proper orchestration of proximity, mobile, and social commerce. The top of Fig.~\ref{fig1} represents the real-world marketing service model, where manufacturers produce and deliver their products through various channels and stores. The bottom of Fig.~\ref{fig1} illustrates the online market of the service model, where users may read other customers' opinions online before deciding to make a purchase. They click on likes, share, and post comments on social events, while retailers may monitor online word-of-mouth (WoM) to sustain a relationship with their customers directly and instantly. In addition, retailers may organize social events to influence and engage consumers' social network friends to participate.

Interestingly, Fig.~\ref{fig1} illustrates that a variety of O2O channels can be applied to create multichannel consumer experiences. This gives rise to \emph{O2O communications}, which comes in different flavors. Cross-channel retailing is defined as the transmission of content through various media in marketing and interaction design. \emph{Omnichannel} retailing is the evolution of cross-channel retailing, putting an emphasis on providing real-time, seamless, consistent, and personalized consumer experiences through all available shopping channels such as social, mobile, store, online-to-offline, and offline-to-online. Overall, O2O communications aim at maximizing the use of both online and offline resources and promoting each other in order to create novel win-win business opportunities~\cite{TWLC15}. 

To better understand how O2O communications may be leveraged for realizing the Internet of No Things, Table~\ref{tab:O2O} compares the advantages and disadvantages of online and offline domains in greater detail. Apart from its various commerce related aspects, the online advantages of particular interest comprise lower costs, wider range of information sources, one-click control simplicity, and superior scalability. By means of O2O communications, these online advantages may be combined with the higher user trust, deeper experiences, instant gratification, and multi-modal physical interactions with people, products, and localized services in the offline domain, treating online and offline channels as \emph{complementary} rather than competitive.

\section{Extended Reality (XR)}
\label{sec:XR}
\begin{figure*}[t]
\centering
\includegraphics[width=.9\textwidth]{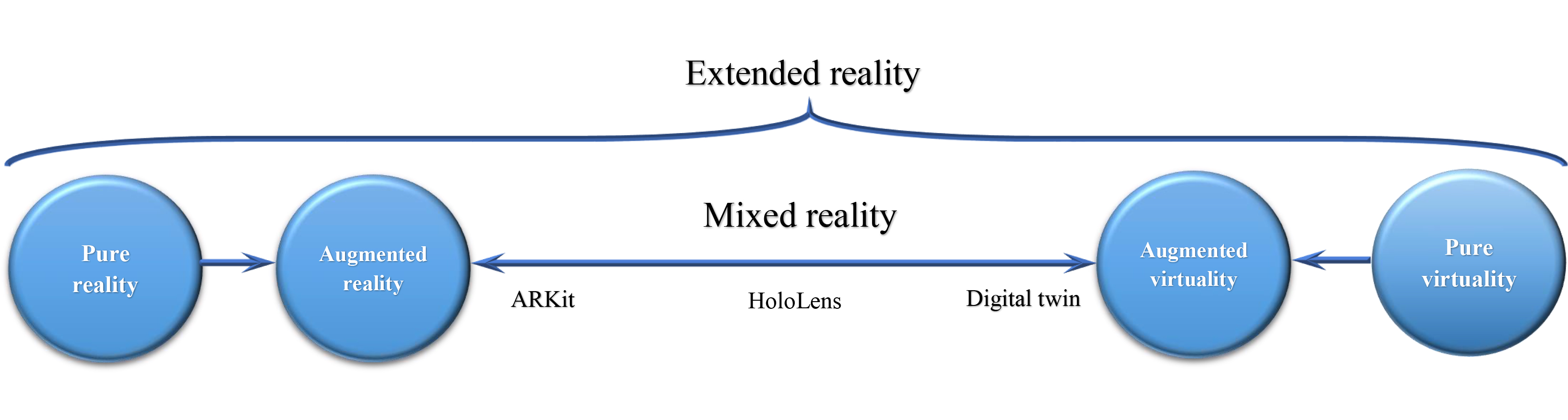}
\caption{The reality-virtuality continuum, ranging from pure reality (offline) to pure virtuality (online).}
\label{fig2}
\end{figure*}
Despite the offline domain's superiority of providing deeper experiences in local environments, \emph{augmented reality (AR)} provides the capabilities to make similar experiences from far away. A good AR example is Shopify's ARKit, which allows smartphones to place 3D models of physical items and see how they would look in real life. To do so, AR overlays virtual objects/images/information on top of a real-world environment. 

AR and the aforementioned VR are part of the so-called reality-virtuality continuum shown in Fig.~\ref{fig2}. The reality-virtuality continuum ranges from pure reality (offline) to pure virtuality (online), as created by VR. Both reality and virtuality may be augmented, leading to AR on one side of the continuum and augmented virtuality (AV) on the other. AR enables the live view of a physical, real-world environment, whose elements are augmented by computer-generated perceptual information, ideally across multiple sensory modalities, including visual, auditory, haptic, somatosensory, and olfactory information. The overlaid sensory information can be constructive (i.e., additive to the natural environment) or destructive (i.e., masking of the natural environment). In doing so, AR alters one's perception of the real-world environment, as opposed to VR, which replaces the real-world environment with a simulated one. Conversely, AV occurs in a virtual environment, where a real object is inserted into a computer-generated environment. An illustrative AV example is an aircraft maintenance engineer, who is able to visualize a real-time model, often referred to as \emph{digital twin}, of an engine that may be thousands of kilometers away (digital twin will be further discussed in Section~\ref{sec:ESP}).

The term \emph{mixed reality (MR)} includes AR, AV, and mixed configurations thereof, blending representations of virtual and real-world elements together in a single user interface. MR helps bridge the gap between real and virtual environments, whereby the difference between AR and AV reduces to where the user interaction takes place. If the interaction happens in the real world, it is considered AR. By contrast, if the interaction occurs in a virtual space, it is considered AV. The flagship MR device is Microsoft's HoloLens, which is the first self-contained, holographic computer that enables users to engage with digital content and interact with holograms in their surrounding environment. 

A recently emerging term is the so-called \emph{extended reality (XR)}, which refers to all real-and-virtual combined environments as well as \emph{human-machine interaction}~\cite{FBGL18}. XR helps improve the time-space flexibility of users by avoiding their need to be at the same place at the same time when working on a project, thus empowering industry with a faster and more powerful decision making process. The areas where most industries apply XR is in remote guidance systems for performing complex tasks such as maintenance and assembly. According to Qualcomm, XR will be the next-generation mobile computing platform that brings the different forms of reality together in order to realize the entire reality-virtuality continuum of Fig.~\ref{fig2} for the extension of human experiences. For illustration, Table~\ref{tab:XR} compares the main characteristics of XR with those of VR, AR, and MR. Note that only XR is suitable to support human-machine interaction as envisioned by the Tactile Internet (see Section~\ref{sec:intro}). Furthermore, the advent of XR technologies pose serious challenges to handheld devices such as smartphones since their design principles focused mainly on ergonomics and battery life. Instead of bearables, it is more suitable to use wearables such as VR and AR head-mounted devices (HMD), while outsourcing their computation and mobility to the mobile edge cloud.
\begin{table}[t]
\centering 
\caption{Comparison of VR, AR, MR, and XR.}
\label{tab:XR}
\begin{tabular}{|l||l|l|l|l|}
\hline
\multicolumn{1}{|c||}{\textbf{Characteristics}} & \textbf{VR} & \textbf{AR} & \textbf{MR} & \textbf{XR} \\ \hline
\hline
Is user aware of real world?                                       & No & Yes & Yes & Yes 
\\ \hline
\hline
Is it possible to create mediated   & Yes & No & Yes & Yes  \\
virtuality? & & & &
\\ \hline
\hline
Is extent of virtuality controllable?    & No & No  & Yes & Yes
 \\ \hline
 \hline
Does it support human-machine inter- & No & No & No & Yes \\
action in real and virtual worlds? & & & & 
\\ \hline
\end{tabular}
\end{table}

Clearly, the discussion above demonstrates the need for designing mechanisms to deal with the heavy traffic loads put on underlying communication networks, given that XR applications have stringent capacity, latency, and reliability requirements to ensure a satisfactory quality-of-immersion (QoI). In fact, according to a recent ABI Research and Qualcomm study~\cite{ABI17}, some of the most exciting AR and VR use cases include remotely controlled devices and the Tactile Internet. More precisely, looking beyond the remote machine control of today to the future Tactile Internet, ultra-low latency feedback below 5 milliseconds will enable novel uses of multisensory remote tactile control with responses rapid enough to be imperceptible to the operator.

\section{Toward the Internet of No Things: Invisible-to-Visible (I2V) Technologies}
\label{sec:I2V}
In this section, we elaborate on how emerging O2O and XR technologies may be leveraged to realize the Internet of No Things. Recall from Section~\ref{sec:intro} that future fully interconnected VR systems will leverage on the growing number of drones, robots, and self-driving vehicles. A very interesting example of future connected-car technologies that merges real and virtual worlds to help drivers `see the invisible' is Nissan's recently unveiled \emph{invisible-to-visible (I2V)} technology concept~\cite{Nissan19}. I2V creates a three-dimensional immersion connected-car experience that is tailored to the driver's interests. More specifically, by merging information from sensors outside and inside the vehicle with data from the cloud, I2V enables the driver and passengers not only to track the vehicle's immediate surroundings but also to anticipate what's ahead, e.g., what's behind a building or around the corner. Although the initial I2V proof-of-concept demonstrator used AR headsets (i.e., wearables), Nissan envisions to turn the windshield of future self-driving cars into a portal to the virtual world, thus finally evolving from wearables to nearables, as discussed in Section~\ref{sec:intro} in the context of the Internet of No Things.

I2V is powered by Nissan's \emph{omnisensing} technology, a platform originally developed by the video gaming company Unity Technologies, which acts as a hub gathering real-time data from the traffic environment and from the vehicle's surroundings and interior to anticipate when people inside the vehicle may need assistance. The technology maps a 360-degree virtual space and gives guidance in an interactive, human-like way, such as through avatars that appear inside the car. It can also connect passengers to people in the so-called \emph{Metaverse} virtual world that is shared with other users. In doing so, people may appear inside the car as AR avatars to provide assistance or company. For instance, when visiting a new place, I2V can search within the Metaverse for a knowledgeable local guide. The information provided by the guide may be stored in the cloud such that others visiting the same area can access it or may be used by the onboard AI system for a more efficient drive through local areas. 

Clearly, I2V opens up endless opportunities by tapping into the virtual world. In fact, the emerging IEEE P1918.1 standard, briefly mentioned in Section~\ref{sec:intro}, highlights several key use cases of the Tactile Internet, including not only the automative control of connected/autonomous driving but also the remote control of robots (i.e., teleoperation). According to~\cite{HaJL19}, the vastly progressing smart wearables such as exoskeletons and VR/AR devices effectively create real-world avatars, i.e., tactile robots connected with human operators via smart wearables, as a central \emph{physical embodiment} of the Tactile Internet. More specifically, the authors of~\cite{HaJL19} argue that the Tactile Internet creates the new paradigm of an immersive coexistence between humans and robots in order to achieve tight physical human-robot interaction (pHRI) and entanglement between man and machine in future locally connected \emph{human-avatar/robot collectives}. Assistive exoskeletons are thereby envisaged to become an important element of the Tactile Internet in that they extend user capabilities or supplement/replace some form of function loss, e.g., lifting heavy objects or rehabilitation systems for people with spinal cord injury. In addition, many studies have shown that the physical presence of robots benefited a variety of social interaction elements such as persuasion, likeability, and trustworthiness. 

\subsection{Extrasensory Perception Network (ESPN)}
In contemporary physics, there exists the so-called ``principle of nonlocality,'' also referred to as \emph{quantum-interconnectedness} of all things by quantum physicists such as David Bohm, which transcends spatial and temporal barriers. Nonlocality occurs due to the phenomenon of entanglement, where a pair of particles have complementary properties when measured. 

Quantum-interconnectedness might be the cause of another phenomenon widely known as \emph{extrasensory perception (ESP)}, which allows humans to have nonlocal experiences in space and time. ESP denotes anomalous processes of information or energy transfer. According to Wikipedia, ESP is also called \emph{sixth sense}, which includes claimed reception of information not gained through the recognized five physical senses, but sensed with the mind. There exist different types of possible ESPs. In this paper, we focus on the following two ESPs to illustrate nonlocal experiences in both space and time: \emph{clairvoyance} (i.e., viewing things or events at remote locations) and \emph{precognition} (i.e., viewing future events before they happen). While clairvoyance may be viewed as the ability to perceive the hidden present, precognition is a \emph{forecast} (not prophecy) of events to come about in the future unless one does something to change them based on the perceived information. In other words, precognition may be viewed as a probable future that does not necessarily have to become actualized, depending on one's behavior. Thus, steering human behavior based on precognition appears to be a powerful means to prevent undesirable future events from happening. 

Note that despite reports based on anecdotal evidence, there has been no convincing  scientific evidence that ESP exists after more than a century of research. However, instead of rejecting ESP as pseudoscience, in this paper we argue that with the emergence of XR it might become possible to disrupt the old impossible/possible boundary and mimic ESP by using multi-/omnichannel O2O communications as an underlying extrasensory perceptual channel. 

Let our point of departure be Joseph A. Paradiso's pioneering work on ESP in an IoT context at MIT Media Lab~\cite{DuPa14}. In a sensor-driven world, network-connected sensors embedded in anything function as extensions of the human nervous system and enable us to enter the long-predicted era of \emph{ubiquitous computing}, as envisioned by Mark Weiser as the ``computer for the 21st century'' more than a quarter of century ago. In addition to visualizing mostly invisible sensor data, Paradiso investigated the impact of ubiquitous computing on the meaning of \emph{presence}. In~\cite{DuPa14}, the authors showed that network-connected sensors and computers make it possible to virtually travel to distant environments and ``be'' there in real time. Interestingly, in their final remarks, the authors concluded that future technologies will fold into our surroundings that help us to get our noses off the smartphone screen and back into our environments, thus making us more (rather than less) present in the world around us.

\begin{figure}[t]
\centering
\includegraphics[width=.5\textwidth]{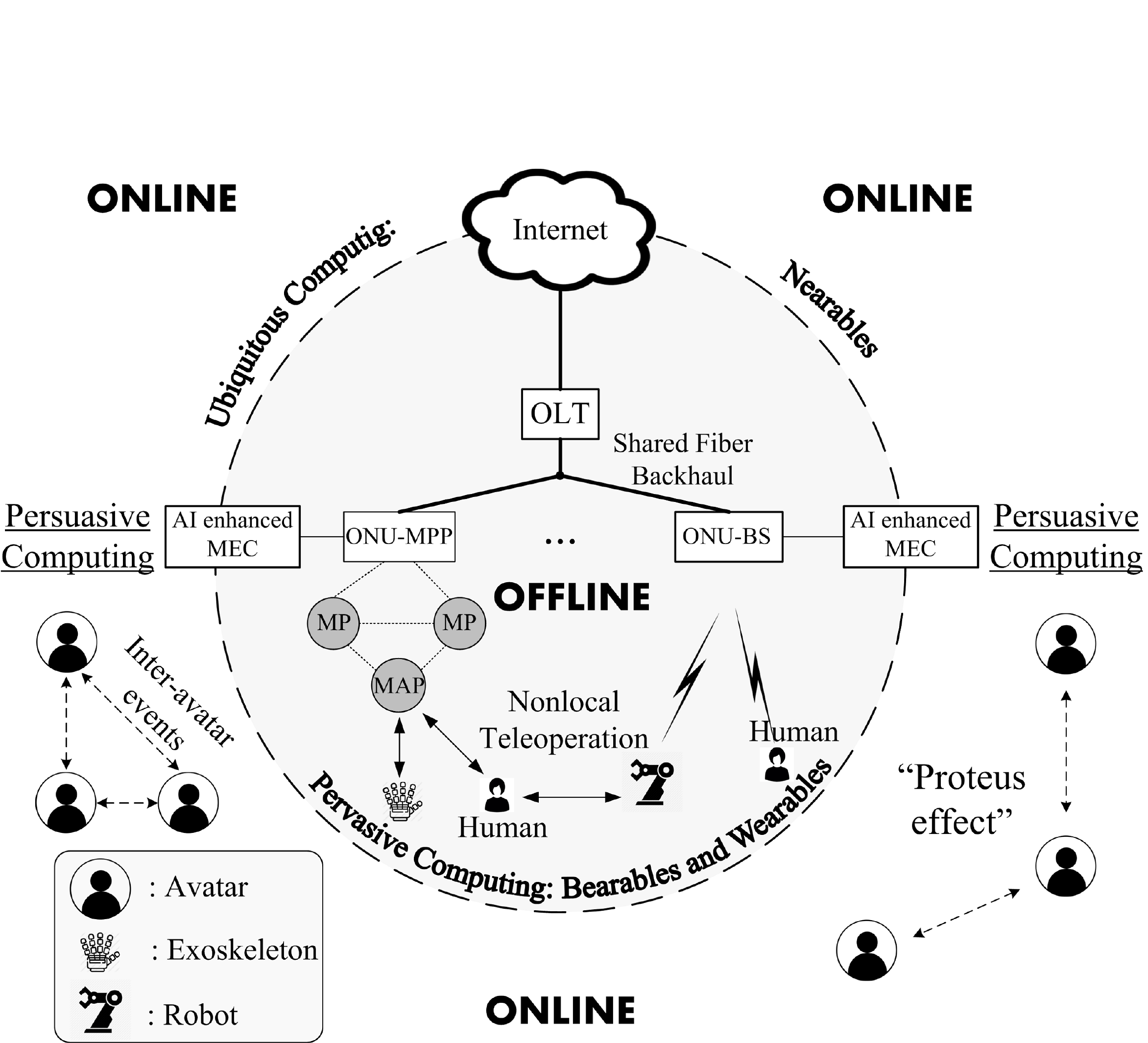}
\caption{Extrasensory perception network (ESPN) architecture integrating ubiquitous, pervasive, and persuasive computing.}
\label{fig3}
\end{figure}
Fig.~\ref{fig3} depicts the architecture of our proposed \emph{ESP network (ESPN)} in greater detail. Its physical network infrastructure consists of a fiber backhaul shared by WLAN mesh portal points (MPPs) and cellular base stations (BSs) that are collocated with optical network units (ONUs). Recently, in~\cite{MaEb19}, we studied the emerging Tactile Internet as one of the most interesting 5G low-latency applications enabling novel immersive experiences by means of haptic communications. Based on real-world haptic traces, we studied the use case of \emph{nonlocal teleoperation} between a human operator and teleoperator robot, which are both physical, i.e., offline, entities (see also Fig.~\ref{fig3}). We showed that AI enhanced MEC helps decouple haptic feedback from the impact of extensive propagation delays by forecasting delayed or lost haptic feedback samples. This enables humans to perceive remote task environments in real-time at a 1-ms granularity. 

As shown in Fig.~\ref{fig3}, our proposed ESPN integrates the following three evolutionary stages of mobile computing: ($i$) ubiquitous, ($ii$) pervasive, and ($iii$) persuasive computing. Ubiquitous computing is embedded in the things surrounding us (i.e., nearables), while pervasive computing involves our bearables and wearables. In the subsequent subsection, we use our developed AI enhanced MEC to realize persuasive computing, which aims at changing the behavior of users through social influence via human-machine interaction utilizing sensors and machine learning algorithms. An interesting phenomenon for changing behavior in an online virtual environment is the so-called ``Proteus effect,'' where the behavior of individuals is shaped by the characteristics and traits of their avatars, especially through interaction during inter-avatar events. 

\subsection{Nonlocal Awareness of Space and Time}
\label{sec:ESP}
In this section, we elaborate on how awareness of nonlocal events in space and time can be realized in our proposed ESPN via O2O communications and XR.

\subsubsection{Clairvoyance}
Clearly, a straightforward approach to realize clairvoyance is the use of network-connected drones, robots, and self-driving vehicles that take cameras to places humans could never imagine reaching, as mentioned in Section~\ref{sec:intro}. Likewise, recall that network-connected sensors allow humans to see obstacles hidden from plain sight. Collecting real-world sensor data feeds and thereby absorbing the real world into the virtual create what David Gelernter calls \emph{mirror worlds}, which give rise to new opportunities for providing user-intended services. For instance, according to Google's chief economist Hal Varian, Google Trends help in ``predicting not the future but the present,'' e.g., the search terms may predict accurately whether a searcher has the flu. Unlike virtual avatars, teleoperated robots may be used for physically \emph{embodied communication}, which was shown to enhance telepresence in a variety of social tasks via multisensory tactile control. 

Typically, there exists a spatial difference between virtual and physical topologies. In our ESPN, this spatial difference can be overcome by sending a \emph{nudge} given by avatar $i$ to intended avatar $j$ to the corresponding MEC server, which then physically nudges the human belonging to avatar $j$ through his wearable(s), e.g., smart jacket. That human in turn may use embodied communication to establish an embodied encounter and proximate relation with the human belonging to avatar $i$.

\subsubsection{Precognition}
In the following, we extend the above virtual/physical nudges to cognitive nudges for achieving precognition in our ESPN. Specifically, we extend our AI enhanced MEC forecasting scheme in~\cite{MaEb19} for using persuasive computing as a tool to explore causal relationships and influence task completion. 

Our AI enhanced MEC forecasting scheme used a type of parameterized artificial neural network (ANN) known as multi-layer perceptron (MLP), which we trained by using haptic traces obtained from real-world teleoperation experiments. It was shown in~\cite{MaEb19} that the developed MLP is able to achieve a high forecasting accuracy (mean squared error below 1\textperthousand) for nonlocal teleoperation. Note that in general the training phase of an ANN is highly critical for setting the weights of its neurons properly. However, the training may become irrelevant in changing or unstructured real-world environments, resulting in a decreased forecasting accuracy. How can we know when or even before this happens?

One way to achieve this is to leverage the aforementioned mirror worlds (e.g., teleoperator robot's digital twin, as discussed in Fig.~\ref{fig2}) for realizing an \emph{omniscient oracle} that uses its context-awareness of the remote task environment for monitoring current teleoperation conditions and detecting faults/failures. In doing so, the effectiveness of our AI enhanced MEC forecasting scheme can be estimated by giving it self-awareness of changing haptic trace based training requirements. To quantify the decreasing effectiveness, the omniscient oracle computes the metric \emph{regret}, which measures the future regret a human operator will have after relying on our haptic sample forecasting scheme even though it has become less trustworthy. Generally, regret is defined as the difference between the reward earned by the policy under study and the best possible reward that could have been obtained by an omniscient oracle. In our teleoperation scenario, we define regret as the difference between the achievable and the optimum task execution times. Note that the metric regret is used to influence the human operator's decision to abort the teleoperation before unintended consequences might occur. 

\begin{figure}[t]
\centering
\includegraphics[width=.45\textwidth]{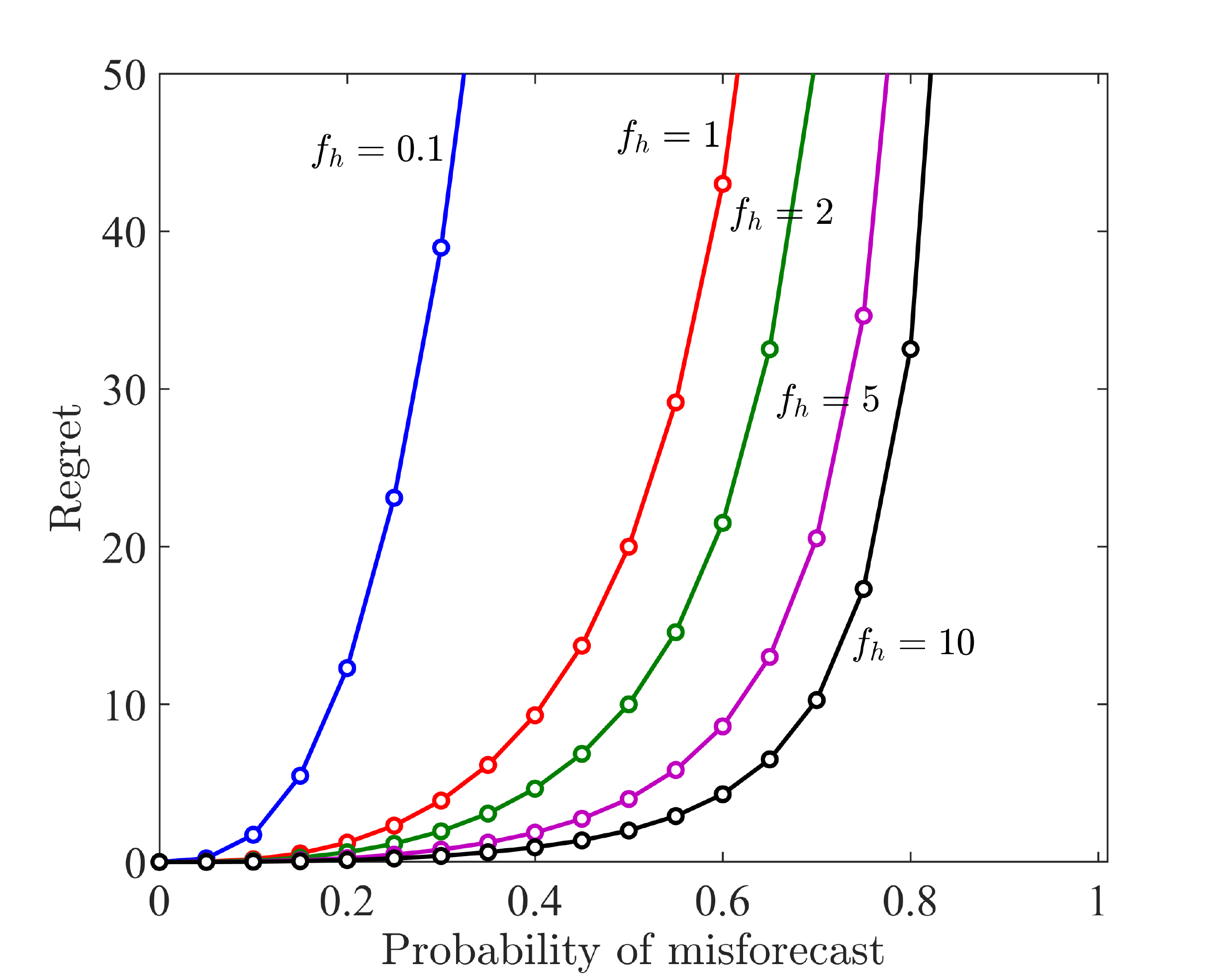}
\caption{Precognition measured as future regret (given in seconds) of relying on MLP-based ANN vs. probability of misforecast with different human capability $f_h$ (given in operations per second) for nonlocal task completion.}
\label{fig4}
\end{figure}
For illustration, Fig.~\ref{fig4} depicts regret as a function of the probability of misforecast, which is defined as the probability that $k$ of the last $n$ received haptic feedback samples are subject to misforecast. Further, let us consider a physical task that requires 20 operations for completion and let $f_h$ denote the number of operations per second the human operator is capable of executing under ideal conditions, i.e., probability of misforecast is zero. Clearly, task completion time increases for increasing probability of misforecast. Fig.~\ref{fig4} shows the general trade-off between the human operator's capability and our MLP-based ANN's forecasting accuracy for $k=3$ and $n=5$, where a less skillful human operator would require a more accurate AI to remain at the same level of regret. More importantly, we observe that especially less skillful human operators with a small $f_h$ may learn from an increasing regret to refrain from blindly relying on a presumably intact AI. 

\section{6G Vision: Putting (Internet of No) Things in Perspective}
\label{sec:6G}
Envisioned 6G mobile networks are anticipated to enable many exciting services and applications that allow for the inclusion of additional human sensory information and even human emotions by making extensive use of powerful knowledge discovery, machine learning, and context awareness techniques. In~\cite{DaBe18}, the authors advocate that 6G should embrace a new mode of thinking from the get-go by including social awareness and understanding the social impact of advanced technologies. They argue that deepened personalization of 6G services that could \emph{predict future events} for the user and provide good advice would certainly be appreciated. Clearly, these ideas are similar to our investigated concept of precognition.

In~\cite{SaBC19}, the authors observed that the ongoing deployment of 5G cellular systems is exposing their inherent limitations compared to the original premise of 5G as an enabler for the \emph{Internet of Everything (IoE)}. They argue that 6G should not only explore more spectrum at high-frequency bands but, more importantly, converge upcoming technological trends. Their bold, forward-looking vision of 6G systems and comprehensive research agenda intend to serve as a basis for stimulating more out-of-the-box research that will drive the 6G revolution. More specifically, they claim that there will be the following four driving applications behind 6G: ($i$) multisensory XR applications, ($ii$) connected robotics and autonomous systems, ($iii$) wireless brain-computer interaction (a subclass of human-machine interaction), as well as ($iv$) blockchain and distributed ledger technologies. Among other 6G driving trends and enabling technologies, they emphasize the importance of haptic and even empathic communications, edge AI, the emergence of smart surfaces/environments and new human-centric service classes, as well as the end of the smartphone era, given that smart wearables are increasingly replacing the functionalities of smartphones. 

We believe that the Internet of No Things with its underlying human-intended services and our proposed ESPN for the nonlocal extension of human ``sixth-sense" experiences in both space and time may serve as a useful stepping stone towards realizing the far-reaching 6G vision above. It's still early in the game. But maybe it turns out that 6G will usher in the Internet of \emph{No Things instead of Everything}. After all, the best things in life aren't things they say.

\section{Conclusions}
\label{sec:conclusions}
We have explored the role of O2O communications and XR in extending the digital and physical capabilities of humans and their experiences in the Internet of No Things, which aims at tying online and offline worlds closer together. We argued that with the emergence of XR it might become possible to disrupt the old impossible/possible boundary and mimic ESP by using multi-/omnichannel O2O communications as an underlying extrasensory perceptual channel. Our proposed ESPN integrates the three evolutionary stages of mobile computing---ubiquitous, pervasive, and persuasive computing---to perceive the hidden present (clairvoyance) and prevent undesirable future events from happening (precognition) based on the nonlocal awareness of space and time. 

At the heart of the experience matrix of today's O2O commerce lies the replacement of the consumption of tangible goods with the consumption of intangible experiences. In their latest book \emph{Capitalism without Capital,} J. Haskel and S. Westlake made the important observation that the rise of the intangible economy may lead to a decreasing openness to experience among certain people, thus exacerbating the causes of increasing inequality in our society. To cope with the intangible inequality, the Internet of No Things has to unleash its full potential by making people more interested and tolerant of new experiences for the benefit of a whole society. 

\section*{Acknowledgment}
The authors are thankful to Saeid Jamshidi for his collaboration.

\begin{IEEEbiographynophoto}{Martin Maier} 
is a full professor with the Institut National de la Recherche Scientifique (INRS), Montr\'eal, Canada. He was educated at the Technical University of Berlin, Germany, and received M.Sc. and Ph.D. degrees (summa cum laude) in 1998 and 2003, respectively. He was a co-recipient of the 2009 IEEE Communications Society Best Tutorial Paper Award. Further, he was a Marie Curie IIF Fellow of the European Commission from March 2014 through February 2015. In March 2017, he received the Friedrich Wilhelm Bessel Research Award from the Alexander von Humboldt (AvH) Foundation in recognition of his accomplishments in research on FiWi enhanced networks. In May 2017, he was named one of the three most promising scientists in the category ``Contribution to a better society" of the Marie Sk\l odowska- Curie Actions (MSCA) 2017 Prize Award of the European Commission. He is the founder and creative director of the Optical Zeitgeist Laboratory (www.zeitgeistlab.ca).
\end{IEEEbiographynophoto}
\vspace{-17.5cm}
\begin{IEEEbiographynophoto}{Amin Ebrahimzadeh}
received his B.Sc. and M.Sc. degrees in Electrical Engineering from the University of Tabriz in 2009 and 2011, respectively. From 2011 to 2015, he was with the Sahand University of Technology (SUT), Iran. He is currently pursuing his Ph.D. in the Optical Zeitgeist Laboratory at INRS, Montr\'eal, Canada. His research interests include fiber-wireless networks, Tactile Internet, teleoperation, artificial intelligence enhanced multi-access edge-computing (MEC), and multi- robot task allocation.
\end{IEEEbiographynophoto}


\begin{thebibliography}{1}
\bibitem{EPBD18}
M. S. Elbamby, C. Perfecto, M. Bennis, and K. Doppler, ``Toward Low-Latency and Ultra-Reliable Virtual Reality,'' \emph{IEEE Network}, vol. 32, no. 2, pp. 78-84, Mar./Apr. 2018.
\bibitem{BBMD17}
E. Ba\c stu\u g, M. Bennis, M. M\' edard, and M. Debbah, ``Toward Interconnected Virtual Reality: Opportunities, Challenges, and Enablers,'' \emph{IEEE Communications Magazine}, vol. 55, no. 6, pp. 110-117, Jun. 2017.
\bibitem{MaEC18}
M. Maier, A. Ebrahimzadeh, and M. Chowdhury, ``The Tactile Internet: Automation or Augmentation of the Human?,'' \emph{IEEE Access}, vol. 6, pp. 41607-41618, Jul. 2018.
\bibitem{AKLY18}
I. Ahmad, T. Kumar, M. Liyanage, M. Ylianttila, T. Koskela, T. Braysy, A. Anttonen, V. Pentikinen, J.-P. Soininen, and J. Huusko, ``Towards Gadget-Free Internet Services: A Roadmap of the Naked world,'' \emph{Elsevier Telematics and Informatics}, vol. 35, no. 1, pp. 82-92, Apr. 2018.
\bibitem{Oren18}
A. Orendorff, ``O2O Commerce: Conquering Online-to-Offline Retail's Trillion Dollar Opportunity,'' May 21, 2018. [Online: Accessed on June 15, 2019]
\bibitem{KiCB17}
W. Kim, S. Chung, and Y. H. Bae, ``O2O Trend and Future: Focused on Difference from Each Case,'' \emph{Journal of Marketing Thought}, vol. 3, no. 4, pp. 42-59, Feb. 2017.
\bibitem{TWLC15}
T.-M. Tsai, W.-N. Wang, Y.-T. Lin, and S.-C. Choub, ``An O2O Commerce Service Framework and its Effectiveness Analysis with Application to Proximity Commerce,'' \emph{Elsevier Procedia Manufacturing}, vol. 3, pp. 3498-3505, 2015.
\bibitem{FBGL18}
\AA. Fast-Berglund, L. Gong, and D. Li, ``Testing and Validating Extended Reality (XR) Technologies in Manufacturing,'' \emph{Procedia Manufacturing}, vol. 25, pp. 31-38, May 2018.
\bibitem{ABI17}
ABI Research and Qualcomm, ``Augmented and Virtual Reality: the First Wave of 5G Killer Apps,'' \emph{White Paper}, Feb. 2017.
\bibitem{Nissan19}
Nissan Newsroom, ``Nissan unveils Invisible-to-Visible technology concept at CES: Future connected-car technology merges real and virtual worlds to help drivers `see the invisible','' Jan. 3, 2019. [Online: Accessed on June 15, 2019]
\bibitem{HaJL19}
S. Haddadin, L. Johannsmeier, and F. D. Ledezma, ``Tactile Robots as a Central Embodiment of the Tactile Internet,'' \emph{Proceedings of the IEEE}, vol. 107, no. 2, pp. 471-487, Feb. 2019.
\bibitem{DuPa14}
G. Dublon and J. A. Paradiso, ``Extra Sensory Perception,'' Scientific American, vol. 311, no. 1, pp. 36-41, Jul. 2014.
\bibitem{MaEb19}
M. Maier and A. Ebrahimzadeh, ``Towards Immersive Tactile Internet Experiences: Low-Latency FiWi Enhanced Mobile Networks With Edge Intelligence [Invited],'' \emph{IEEE/OSA Journal of Optical Communications and Networking, Special Issue on Latency in Edge Optical Networks}, vol. 11, no. 4, pp. B10-B25, Apr. 2019.
\bibitem{DaBe18}
K. David and H. Berndt, ``6G Vision and Requirements: Is There Any Need for Beyond 5G?,'' \emph{IEEE Vehicular Technology Magazine}, vol. 13, no. 3, pp. 72-80, Sep. 2018.
\bibitem{SaBC19}
W. Saad, M. Bennis, and M. Chen, ``A Vision of 6G Wireless Systems: Applications, Trends, Technologies, and Open Research Problems,'' \emph{arXivorg}, Cornell University, Ithaca, NY 14850, USA, Feb. 2019.

\newpage

\end{thebibliography}
\end{document}